# Monte Carlo phonon transport simulations in hierarchically disordered silicon nanostructures

Dhritiman Chakraborty[*], Samuel Foster, and Neophytos Neophytou

[1]School of Engineering, University of Warwick, Coventry, CV4 7AL, UK
[*] D.Chakraborty@warwick.ac.uk

## Abstract

Hierarchical material nanostructuring is considered to be a very promising direction for high performance thermoelectric materials. In this work we investigate thermal transport in hierarchically nanostructured silicon. We consider the combined presence of nanocrystallinity and nanopores, arranged under both ordered and randomized positions and sizes, by solving the Boltzmann transport equation using the Monte Carlo method. We show that nanocrystalline boundaries degrade the thermal conductivity more drastically when the average grain size becomes smaller than the average phonon mean-free-path. The introduction of pores degrades the thermal conductivity even further. Its effect, however, is significantly more severe when the pore sizes and positions are randomized, as randomization results in regions of higher porosity along the phonon transport direction, which introduce significant thermal resistance. We show that randomization acts as a large increase in the overall effective porosity. Using our simulations, we show that existing compact nanocrystalline and nanoporous theoretical models describe thermal conductivity accurately under uniform nanostructured conditions, but overestimate it in randomized geometries. We propose extensions to these models that accurately predict the thermal conductivity of randomized nanoporous materials based solely on a few geometrical features. Finally, we show that the new compact models introduced can be used within Matthiessen's rule to combine scattering from different geometrical features within ~10% accuracy.

**Keywords:** thermal conductivity, disordered nanomaterials, thermoelectrics; nanotechnology; nanoporous materials; Monte Carlo simulations.

# I. Introduction

Highly disordered nanostructures are one of the most promising ways to achieve very high thermoelectric (TE) efficiencies and, thus, engineering such materials has recently attracted significant attention. Strong disorder, and more specifically disorder on hierarchical length scales, originating from various types of defects, can scatter phonons of different wavelengths throughout the spectrum and drastically reduce thermal conductivity. This approach substantially improves thermoelectric efficiency and it is currently being employed in a variety of new generation thermoelectric materials. For example, using hierarchical inclusions at the atomic scale, the nanoscale, and the mesoscale in the PbTe – SrTe system, Biswas *et* al. reported a lattice thermal conductivity $\kappa$ of 0.9 Wm$^{-1}$ K$^{-1}$ at 915 K and a *ZT* of 2.2.[1] More recently, using this method for the p-type Pb$_{0.98}$Na$_{0.02}$Te-SrTe system, Tan *et* al. reported an even lower $\kappa$ of 0.5 W K$^{-1}$m$^{-1}$ and a higher *ZT* of 2.5 at 923K.[2] Reports also show that hierarchical nanostructures can improve the thermoelectric power factor as well.[3,4,5,6]

Specifically, for Si-based materials, Si nanowires have been reported to exhibit thermal conductivities close to, or even below the amorphous limit ($\kappa$ < 2 Wm$^{-1}$ K$^{-1}$), which allowed a 50-fold increase in *ZT* to *ZT*~0.5 by surface roughness engineering.[7,8,9] (The experimentally determined $\kappa$ of amorphous silicon thin films is in the range of 1 - 2 Wm$^{-1}$ K$^{-1}$ at room temperature).[10] Similar observations have been reported for SiGe nanowires[11] and silicon thin films of 2 - 6 nm in thickness.[12,13] Drastic reductions in thermal conductivity were also reported in nanocrystalline materials. Wang *et* al.[14] showed that the room temperature silicon thermal conductivity decreases from 81 Wm$^{-1}$ K$^{-1}$ to 24 Wm$^{-1}$ K$^{-1}$ as the average grain size decreases from 550 nm to 76 nm, whereas $\kappa$ below 5 Wm$^{-1}$ K$^{-1}$ has been reported for average grain sizes of about 10 nm.[15] For grain sizes of 3 nm Nakamura *et* al. reported $\kappa$ = 0.787 ± 0.12 Wm$^{-1}$ K$^{-1}$.[16,17]

In addition, single-crystalline silicon membranes with nanoscopic pores exhibit reproducibly low $\kappa$ around 1-2 Wm$^{-1}$ K$^{-1}$,[18,19,20] while still maintaining sufficient electronic properties. Nanostructures that combine the effects of alloying, nanocrystallinity, and porosity have started to appear as well, as a means to achieve an even lower $\kappa$. Specifically for the nanocrystalline/nanoporous geometry, a recent Si-based work reported $\kappa$ of 20.8 ± 3.7 Wm$^{-1}$ K$^{-1}$ for an average pore size of ~ 30 nm and grain sizes between 50 and 80 nm.[21] By reducing both pore and grain sizes, however, Basu *et al.* reported $\kappa$ = 1.2 Wm$^{-1}$ K$^{-1}$ at 40% porosity in p-type silicon.[22] A recent work in SiGe nanomeshes, reported ultralow $\kappa$ of 0.55 ± 0.10 Wm$^{-1}$ K$^{-1}$ for SiGe nanocrystalline nanoporous structures, a value well below the amorphous limit.[23]

A significant amount of work can be found in the literature attempting to clarify these experimental observations. However, theoretical investigations of thermal conductivity in highly/hierarchically disordered nanostructures (which include not only crystalline boundaries, but also pores of random sizes placed at random positions) are very limited. Understanding the qualitative and quantitative details of such geometries on the thermal conductivity would allow the design of more efficient thermoelectrics and heat management materials in general. In this work, we solve the Boltzmann transport equation for phonons in disordered Si nanostructures using the Monte Carlo (MC) method. Monte Carlo, which can capture the details of geometry with relative accuracy, is widely employed to understand phonon transport in various nanostructures such as nanowires,[24,25,26] thin films,[27,28] nanoporous materials,[29,30,31,32,33] polycrystalline materials,[10,15,34,35,36] nanocomposites,[37,38] corrugated structures,[39,40,41,42] silicon-on-insulator devices,[43] etc. We consider geometries that include grain boundaries, surfaces, and pores as in realistic nanocomposite materials, which all contribute to reducing thermal conductivity. We show that the influence of randomization in the disorder can have a crucial effect in determining thermal conductivity, despite being usually overlooked. After examining the influence of nanocrystallinity and porosity individually and combined, we validate the simplified compact models commonly employed in the literature. We then propose more accurate models based on simple geometrical configurations that describe the randomization of disorder. These improved models could serve as a valuable tool for materials design and for experimentalists to more accurately evaluate a first order interpretation of their results, without the need of large scale simulations.

The paper is organized as follows: In Sec. II we describe our theoretical and computational approach. In Sec. III we present our results on the effects of disorder (nanocrystallinity, nanoporosity, and both combined) on thermal conductivity $\kappa$. In Sec. IV we validate existing analytical models for such geometries and develop our more accurate compact models. Finally, in Sec. V we conclude.

## II.   Approach

The Monte Carlo (MC) approach has been adopted for a semiclassical particle based description of phonon transport. For computational efficiency we consider a two-dimensional (2D) simulation domain of length $L_x$ = 1000 nm and width $L_y$ = 500 nm. The domain is populated with nanostructured features as shown in Fig. 1. The MC simulation method is described adequately in the literature, but because our method differs in some details, below we describe our numerical scheme. We use the "single-phonon MC" approach which differs from the multi-phonon MC approach described in various works in the literature[24,44,45,46,47] in terms of phonon attributes book-keeping. In a multi-phonon approach, a number of phonons are initialized simultaneously. Phonon paths, energy and temperature of all cells are traced simultaneously at every time step, and often periodic boundary conditions are employed to remove the effect of the limited simulation domain. In the single phonon approach one phonon is simulated at a time from the domain boundary/edge and propagates through the simulation domain until it exits at either edge. Once the phonon exits, the next phonon is then initialized.

The simulation procedure is then as follows: Phonons enter from either direction of the simulation domain and alternate between free flight and scattering events. The time a phonon spends in the simulation domain until it exists again, is recorded as its time of flight (TOF). The regions at the left/right of the simulation domain are given "hot" and "cold" temperatures, $T_H$ and $T_C$, respectively. The rest of the domain is initially set at the average temperature of $T_H$ and $T_C$ – we label that as $T_{BODY}$. At room temperature, a $\Delta T$ = 20 K is adequate to gather the necessary statistics for simulation convergence, and low enough to ensure the simulation is still within the linear response regime.[31,48]

Phonons are initialized in the contacts only, based on polarization, frequency, velocity, and energy. Phonon probabilities are drawn from a dispersion relation $\omega(q)$, weighted by the Bose Einstein distribution at the given temperature. We use the dispersion relation $\omega(q)$ and corresponding group velocities $v_g(q)$ as described by Pop et al.[46] in Eqs. 1 and 2 below:

$$\omega(q) = v_s + cq^2 \tag{1}$$

$$v_g = \frac{d\omega}{dq} \tag{2}$$

where $q$ is the wave vector norm and $v_s$ and c are fitting parameters to match the thermal conductivity of bulk Si in the [100] direction. The dispersion coefficients we use are $v_s$ = 9.01 × 10³ ms⁻¹ and c = -2 × 10⁻⁷ m²s⁻¹ for the longitudinal acoustic (LA) branch, and $v_s$ = 5.23 × 10³ ms⁻¹ and c = -2.26 × 10⁻⁷ m²s⁻¹ for the transverse acoustic (TA) branches.[48] Following common practice, the contribution of optical phonons is neglected as they have low group velocities and do not contribute significantly to phonon transport[24,46,47,49] (although they indirectly could influence the interaction between optical and acoustic phonons and alter the effective relaxation rates of the acoustic phonons[50]).

Phonons in the simulation domain either scatter, or are in free flight. During free flight, the position $r$ at time $t$ of the phonon is given by the equation:

$$r(t_i) = r(t_{i-1}) + v_g \Delta t \tag{3}$$

Scattering of phonons is caused either by interaction with geometrical features, or by three-phonon internal scattering (Umklapp processes). The three-phonon scattering, which is responsible for the change in the temperature of the domain, is computed in the relaxation time approximation and is a function of temperature and frequency, as:

$$\tau_{TA,U}^{-1} = \begin{cases} 0 & \text{for } \omega < \omega_{1/2} \\ \dfrac{B_U^{TA}\omega^2}{\sinh\left(\dfrac{\hbar}{k_B T}\right)} & \text{for } \omega > \omega_{1/2} \end{cases} \tag{4}$$

where $\omega$ is the frequency, $T$ the temperature, $B_U^{TA}$ = 5.5 × 10⁻¹⁸ s, and $\omega_{1/2}$ is the frequency corresponding to $q = q_{max}/2$. These equations and parameters are well-established and often used to describe relaxation time in phonon Monte Carlo simulations for Si.[24,31,44,45,46,48] Three phonon scattering causes a change in the energy, and thus the temperature ($T$) of the simulation domain "cell" where scattering took place (we use a 1 nm domain discretization). Every time this happens the "cell" temperature either rises or falls. The link between energy and temperature is given by:

$$E = \frac{V}{W} \sum_p \sum_i \left( \frac{\hbar}{\exp\left(\frac{\hbar}{k_i T}\right) - 1} \right) g_i D(\omega_i, p) \Delta\omega \tag{5}$$

where $\omega$ is the frequency, $T$ the temperature, $D$ is the density of states at given frequency and branch polarization, and $g_i$ the polarization branch degeneracy, and V is the volume of the "cell". The temperature of each "cell" is then numerically determined by back-solving Eq. 5 iteratively using the Newton Raphson method. The dissipation/absorption of energy from each 'cell' in this way establishes a temperature gradient under a continuous flow of phonons (shown in Fig. 1a with the yellow to green color scheme). A scaling factor (W = $4 \times 10^5$) is also introduced to scale the number of phonons simulated to the real population of phonons from $6 \times 10^{10}$ μm$^{-3}$ that are present at 300 K for computational efficiency.[31] We keep W constant in the entire simulation domain, where the average temperature is 300 K.[31] Initially, phonons are injected from both ends of the domain at their respective junction temperatures to establish a temperature gradient across the device as seen by the coloring in Fig. 1. An average of 2.5 million phonons are simulated for this, injected from each side.

Once the thermal gradient has converged, another 2.5 million phonons are then injected into the domain from each side. They can make it to the other side, or backscatter to where they originated from. The total energy entering and leaving the simulation domain is calculated by the net sum of the corresponding phonon energies that enter/exit at the hot and cold junctions as specified by Eq. 5. We label the total incident energy from the hot junction as $E_{in}^H$ and the total energy of phonons leaving the simulation domain from the hot junction as $E_{out}^H$. Similarly, $E_{in}^C$ and $E_{out}^C$ are the incoming and outgoing energies at the cold junction. We then determine the average phonon energy flux in the system as:

$$\Phi = \frac{\left(E_{in}^H - E_{out}^H\right) - \left(E_{in}^C - E_{out}^C\right)}{n \langle TOF \rangle} \tag{6}$$

where $n$ is the total number of phonons simulated and $\langle TOF \rangle$ is their average time-of-flight. The simulated thermal conductivity is then extracted as:

$$\kappa_s = \Phi \frac{L_X}{A_C \Delta T} \tag{7}$$

where $A_C$ is the effective (scaled) cross section area of the simulation domain, which together with the scaling factor W above, are used to convert the simulated energy flux to thermal conductivity with the proper units, as well as calibrate our 2D simulation result to the pristine Si bulk thermal conductivity value (details to follow).

Next, to account for the fact that the length of the simulated domain ($L_x$) is smaller than some phonon mean-free-paths, especially at lower phonon frequencies, a scaling of the simulated thermal conductivity ($\kappa_s$) is needed to compute the final thermal conductivity $\kappa$ as[51]:

$$\kappa = \kappa_s \frac{(L_X + \lambda_{pp})}{(L_X)} \tag{8}$$

where $\lambda_{pp}$ is the average phonon mean-free-path (mfp) of Si. Values for the average bulk mfp of phonons in Si at room temperature vary in the literature. In experimental studies, values from 100 nm[52] to 300 nm[53,54] are mentioned (the latter[54] is a study on Si films). In theoretical works an even greater variation – from 43 nm[55], 100 nm[56], 135 nm[51, 57] to 200-300 nm[44] have been reported. Here we chose to use $\lambda_{pp}$ = 135 nm from Jeong et al[51], because in that work the mfp is reported over a variety of temperatures, and we could then validate our scaling method over the entire temperature range. In this way, the finite size of the simulation domain is overcome using the average mean-free-path to scale the simulated thermal conductivity to the actual thermal conductivity of an infinite channel length. This scaling is important for the pristine bulk case of silicon where a large number of phonons have mean-free-paths larger than the simulation domain size, and replaces the need for periodic boundary conditions.[31] It is particularly important in the low temperature range where the low temperature $\kappa$ peak of silicon is observed only after this scaling. It allows us to simulate shorter channels (in the micrometer range), which simplifies the simulation considerably. Thus, with the width of our simulation domain fixed at 500 nm and a scaling factor of W = $4\times10^5$ as specified above,[58] which accounts for the reduction of the number of simulated phonons per unit volume, when using the thickness of 0.1 nm which corresponds to a single atomic layer[59] (to compute the volume V of the 'cells' above), our simulated thermal conductivity is $\kappa_s$ = 130 Wm$^{-1}$K$^{-1}$ at 300K. After using Eq. 8 to account for the finite simulation domain we obtain the pristine bulk silicon $\kappa \sim$ 148 Wm$^{-1}$K$^{-1}$, which is the value for pristine bulk Si at room temperature. [28,48] Note that in the case of nanocrystalline and nanoporous structures (the focus of this work), where the scattering length is determined by the grain sizes and pore distances, this scaling is less important. Indeed, the difference between the calculated thermal conductivity if the mfp scaling is performed using using $\lambda_{pp}$ = 135 nm, vs $\lambda_{pp}$ = 300 nm is at most 15% in the pristine material case, but drops to ~6% in the case where pores are introduced, and becomes insignificant when nanocrystallinity is introduced as well. This is also reinforced by the fact that we have fully diffusive transport in our disordered systems, verified by simulations of channels with different lengths and extraction of the average phonon paths. (See the Appendix for validation of this method and the above statement for different $\lambda_{pp}$ considerations, as well as demonstration of diffusive transport in the channels we simulate).

Thus, scaling by an "effective" thickness we can calibrate the pristine material to Si bulk values, and by scaling with the mean-free-path in Eq. 8 we make it possible to simulate shorter channels and avoid periodic boundary conditions (see Appendix for validation this statement in channels with different lengths). Also, the use of the 'single-phonon' method is computationally simpler since we do not keep track of all phonon positions at the same time. All this simplifies the computation significantly. In addition, although Monte Carlo can be efficient for complex geometries in three dimensions and in the past some of us and other authors had published studies for 3D MC as well,[24,28,31,45,48,49] here we effectively simulate a 2D material, i.e. corresponding to ribbons. This is adequate for exploring the influence of disorder variability (our primary goal). We executed over 1000 simulations, each simulation taking ~8-10h, which is an order of magnitude less computationally expensive compared to 3D simulations.

We consider transport in different nanostructured geometries. In the first case we consider nanocrystalline geometries as shown in Fig. 1b, where the average grain size in the simulation domain defined as

$$<d> = L_x / <N_G> \qquad (9)$$

where $L_x$ is the length of the domain in the transport direction and $<N_G>$ is the average number of grains encountered in that length. Grains in the nanocrystalline case are generated using voronoi tessellations, where grain boundaries are created by considering input values for the number of "seeding points" and the dimensions of the domain.[35] In these structures, thermal conductivity is impeded in two ways - the scattering of phonons due to the grain boundaries and internal three-phonon Umklapp scattering inside the grains. If a phonon reaches a boundary, then a decision is made whether the phonon will transmit to the other side, or reflect. This decision is made upon a probability distribution, which depends on the phonon wave vector, the roughness of the boundary $\Delta_{rms}$ and the angle of incidence between the phonon path and the normal to the grain boundary $\theta_{GB}$ (see Fig. 1f). The transmission probability upon boundary scattering is then given by the commonly employed relation[35]:

$$t_{scatter} = \exp\left(-4q^2 \Delta_{rms}^2 \sin^2\theta_{GB}\right) \quad (10)$$

If the phonon makes it to the other side of the boundary, it continues its path intact. If it is reflected, then another random number, that depends on the specularity parameter $p$ (roughness strength), dictates whether it will scatter specularly, or diffusively.[45] We do not assume that the phonon changes its energy at the interface as is common practice, but only its direction. $p$ takes values from 0 to 1, with $p = 0$ indicating diffusive, randomized reflection angle and $p = 1$ specular reflection where the angle of incidence is the same as the angle of reflection (see Fig. 1f). Using $p$ to determine the specularity of reflection is also applied for pore boundaries – in that case a phonon can only reflect (see Fig. 1e). Specifically in the case of specular pore boundary scattering, the angle the phonon will reflected into, is defined based on geometrical considerations (the angle of incidence is the same as the angle of reflection) as:

$$\theta_{ref} = 2\gamma - \theta_{inc} \quad (11)$$

where $\theta_{inc}$ is the angle of propagation of the phonon relative to the x-axis, and $\gamma$ is the angle formed by the line perpendicular to the pore at the point of interaction and the x-axis, as explained in Fig. 1e.

Note that instead of a constant specularity $p$, in Monte Carlo it is also customary to determine the actual specularity for each phonon using the expression $p(q) = \exp(-4q^2 \Delta_{rms}^2)$, which also allows wave-vector dependence reflections. In that case, what is constant is the surface roughness ($\Delta_{rms}$). Here we use a constant $p$, and the rationale behind studies which use constant $\Delta_{rms}$ versus constant $p$[31,48,60,61,62,63,64] is that the microscopic details of phonon scattering at interfaces are poorly understood anyway.[64] However, either way gives vary similar results without any qualitative or quantitative differences in disordered structures. For example, one can map a specific $\Delta_{rms}$ to a specific specularity; the $p = 0.1$ case in our results below corresponds to $\Delta_{rms} \sim 0.3$ nm (see Appendix), which corresponds well to rough silicon surfaces.[65,66]

## III. Results

Initially simulations were carried out to compare and validate the simulator for bulk values of silicon thermal conductivity. All validation of the simulator was carried out using a fixed simulation domain of length $L_x = 1000$ nm and width $L_y = 500$ nm. Good agreement is found between our simulated results and literature values of silicon thermal conductivity across a large temperature range with several works in the literature. After bulk-Si validation, we proceeded with the analysis of nanostructuring on the thermal conductivity.

<u>Nanocrystalline geometries – influence of grain size and boundary roughness:</u> We begin our investigation with the effects the grain size <$d$> and boundary roughness ($\Delta_{rms}$) the thermal conductivity. The results are shown in Fig. 2 where $\kappa$ is plotted as a function of average grain size <$d$>. Each point is an average of 50 simulations of different geometry realizations. We consider average grain size from <$d$> = 1000 nm down to <$d$> = 50 nm as indicated by the geometry subfigures above the graph in Fig. 2 (from subfigure 1 in the geometry panel where <$d$> = 50 nm to subfigure 6 where <$d$> = 225 nm). Three different values of $\Delta_{rms}$ = 0.25 nm, 1 nm, and 2 nm were simulated, shown in Fig. 2 by the red, blue, and black lines, respectively. Decreasing grain size causes a reduction in $\kappa$, from 97.8 Wm$^{-1}$K$^{-1}$ to 19.9 Wm$^{-1}$K$^{-1}$. This is consistent with other available theoretical[10,14,15,67,68,69,70,71,73,73] and experimental results,[14,21,68,73,74] as shown in the inset in

Fig. 2. Note that our large grain thermal conductivity does not reach the bulk value ($\kappa \sim 140$ Wm$^{-1}$K$^{-1}$) because we consider boundary scattering on the surfaces of the simulation domain. An important observation is that a rapid drop in $\kappa$ is observed for structures in which the average grain size is below the average phonon mean-free-path ($\lambda_{pp} = 135$ nm). For these structures grain boundary scattering has a more dominant role than intrinsic three-phonon scattering. This observation is consistent for the different values of grain boundary roughness.

On the other hand, changes in the values of grain boundary roughness ($\Delta_{rms}$) seem to play a comparatively smaller role in decreasing $\kappa$ (comparing the red, blue, and black lines respectively, in Fig. 2). Phonon paths are already randomized by the numerous grains and intrinsic scattering, and thus the additional randomness from grain boundary roughness plays a minimal role. Similarly, it is also noticeable that as the grain size decreases the variability in the results (the average of the 50 simulations for each point), as indicated by the error bars, also decreases, especially for grain sizes smaller than the mean-free-path.

<u>Nanoporous geometries – influence of porosity and pore roughness:</u>

Figure 3 summarizes the effects of porosity ($\phi$) and pore boundary roughness on the thermal conductivity of nanoporous silicon. Examples of the typical geometries considered, with porosities of 5%, 10% and 15%, for both ordered and disordered configurations, are shown as shown in the geometry panel above Fig. 3. In all panels the channel dimensions are length $L_x = 1000$ nm and width $L_y = 500$ nm. In the ordered geometry cases the pore diameter is fixed at $D = 50$ nm. In the random cases the pores are arranged in random positions and their diameters vary from 10 to 50 nm using a uniform distribution. Here, $\kappa$ is plotted as a function of porosity $\phi$ (x-axis), and results for structures with boundary specularity parameters $p = 1$ (fully specular case, blue line), $p = 0.5$ (green line), and $p = 0.1$ (almost fully diffusive case, red line) are shown. The ordered pore cases are shown in solid and the randomized pore cases in dashed lines. Again, each point is the average of 50 different configurations with the variation bar denoted. Note that for the $p = 1$ case (blue solid line), the boundaries everywhere are completely specular, and for zero porosity, the value of $\kappa$ approaches the bulk 148 Wm$^{-1}$ K$^{-1}$.

Phonons back-scatter on the pores since the pores are large and transmission is not allowed, unlike in the case of grain boundary scattering where transmission of phonons through the grain boundary is statistically allowed. First, we observe that reducing specularity causes a reduction in $\kappa$. However, an order of magnitude decrease in $p$ from $p = 1$ (blue line) to $p = 0.1$ (red line) causes only a ~33% drop in $\kappa$ at most, and more noticeably at low porosities where phonon trajectories might still not be completely randomized.[48] Increasing porosity, on the other hand, causes a significant decrease in $\kappa$ as also observed in previous theoretical[29,31,48,56,75,76,77] and experimental results.[11,18,19,20] In fact, the effect of porosity has a much greater impact than pore roughness. An order of magnitude reduction in specularity causes roughly the same effect as 15 % porosity in the ordered case (blue solid line), an observation consistent with previous works.[31, 48] We note here that Monte Carlo does not account for coherent phonon effects which could affect the phonon spectrum and thermal conductivity, but there is increasing evidence that such effects are important only at low temperatures and weak roughness conditions,[60, 72] whereas here we deal with room temperature and mostly diffusive boundaries.

Nanoporous geometries - influence of *randomized* pore positions and diameters: We next consider the effects of random diameter and pore positions at different porosities as shown in the "random" panels in Fig. 3. A further decrease in thermal conductivity is observed as a consequence of disorder, irrespective of pore boundary specularity. For the diffusive pore case the reduction can vary from ~35% (low porosity) to ~65% (high porosity), which is quite significant, as shown in the inset of Fig. 3. We discuss the reasons behind this in detail in Section IV below when we construct an analytical model to account for this reduction. On the other hand, the influence of specularity in the randomized pore cases is again comparatively minimal and diminishes as porosity increases (blue, green and red dashed lines in Fig. 3).

It is illustrative to separate the two effects that constitute the randomization in the polydispersed nanoporous geometries, i.e. the randomization in pore position and randomization in pore diameter. Figure 4 shows the thermal conductivity if structures with ordered pores and polydispersed pores (solid and dashed red lines – the same as the $p = 0.1$ cases in Fig. 3), and the corresponding thermal conductivity of the structures in which only the pore positions are randomized. Typical geometries are depicted in the schematics of Fig. 4. Clearly, the randomization of the positions alone has a significant effect in lowering the thermal conductivity. It seems that for lower porosities it is the dominant factor for the deviation between the ordered and the polydispersed geometries. For higher porosities both randomized location and randomized diameters have similar influence in further reducing the thermal conductivity from the ordered pore case.

Hierarchical disordered nanostructures - combining nanocrystallinity and porosity: We next combine the effects of nanocrystallinity and porosity, as in realistic nanoporous materials. Again, ordered and randomized pores are considered as shown in Figs. 5a and 5b, respectively, shown here for 5% porosity. Figure 4c plots the thermal conductivity versus the average grain size ($<d>$) for structures with different porosity values ($\phi$). The roughness on the transmittable grain boundaries is fixed to $\Delta_{rms}$ = 1 nm, while for the outer top/bottom boundaries of the simulation domain and the pore boundaries we use a specularity parameter of $p$ = 0.1. Both conditions correspond to rough, almost fully diffusive cases. Again each point shown in Fig. 5c is an average of 50 simulations.

The top blue line depicts the zero porosity case, the same as the initial results for $\Delta_{rms}$ = 1 nm shown in Fig. 2 (blue line). Adding pores in an ordered fashion further reduces $\kappa$. This can be seen for 5% (magenta line), 10% (light blue) and 15% (red) ordered porosity. The thermal conductivity decreases as either porosity increases, or the average grain size $<d>$ decreases, with large porosity dominating at large grain sizes, whereas boundary scattering dominates at small grain sizes. With regard to the variation bars, as porosity increases and/or grain size decreases, scattering becomes more and more randomized, and variations in the thermal conductivity are reduced, as also observed above.

Hierarchical disordered nanostructures - nanocrystallinity with randomized pores: The red-dashed line in Fig. 5c shows the thermal conductivity versus average grain size in the case of a $\phi$ = 15% randomized porous structure. The pores are randomized in terms of diameter and position as indicated in Fig. 5b. The pore sizes are again varied from $D$ = 10 nm to 50 nm in a random fashion using a uniform distribution. As in the case of only pores geometries earlier, randomization in the pore features reduces $\kappa$ significantly. In this case, at $<d>$ = 1000 nm at the right side of Fig. 5c, there is an initial 50% drop in $\kappa$ in the randomized case compared to the ordered (red lines), followed by a slow rate of decrease in $\kappa$ as the grain size decreases. This suggests that a high degree of randomization and small average pore size, makes phonon scattering on pores much more dominant than the intrinsic three phonon scattering and grain boundary scattering. When the grain size becomes very small (below $\lambda_{pp}$ ), then it starts to play an important role again.

## IV. Analytical models – extensions and validation

There are many analytical models available in the literature that describe the effects of material geometry on $\kappa$, either in the presence of grain boundaries,[71,72,78,79] or pores.[75,76,80,81,82] These are based on simple geometrical considerations, and assume uniformity of the corresponding features, but in the case of non-uniformities, or in the presence of two or more types of nanosized features, their accuracy fades. Here we compare our full simulation results to some of these widely employed analytical models found in the literature. We aim to quantify their validity and further develop more accurate models that can capture the effects of non-uniformity in nanostructuring, based again on simple geometrical considerations.

<u>Analytical models - Nanocrystalline case:</u> The analytical models widely employed for nanocrystalline materials, are based on the simple logic that: i) phonons in a nanostructured material undergo additional scattering events at a rate at which they meet the boundaries as they propagate in the material, ii) an additional interface resistance (Kapitza resistance) is introduced due to disruptions in the phonon flow. Based on these principles, a few examples of the form that these models take are given in the works of Nan et al.[78], Yang et al.[79] and Dong et al.[71], given by Eqs. 12- 14, respectively:

$$\kappa = \frac{\kappa_0}{\left[1+\frac{R_K \kappa_0}{d}\right]} \quad (12)$$

$$\kappa = \frac{\kappa_0}{\left[1+\frac{2R_K \kappa_0}{d}\right]} \quad (13)$$

$$\kappa = \frac{\dfrac{\kappa_0}{1+\dfrac{\lambda_{pp}}{d^{\alpha}}}}{\left[1+\dfrac{R_K \kappa_0}{\left(1+\dfrac{\lambda_{pp}}{d^{\alpha}}\right)}\right]} \tag{14}$$

Above, $\kappa_0$ is the bulk thermal conductivity of silicon and $\lambda_{pp}$ is the average phonon-phonon mean-free-path (here $\lambda_{pp}$ = 135 nm), $R_K$ is the Kapitza resistance, $d$ the average grain size (<$d$> in Monte Carlo), and α is a commonly used fitting parameter.[71] Here we use $R_K$ = 1.06 ×10$^9$ Km$^2$W$^{-1}$.[83] Literature values for $R_K$ vary slightly in the range of 1-1.16 × 10$^9$ Km$^2$W$^{-1}$,[71,83,84,85,86,87,88] a variation that makes only very little qualitative difference to the results we show below (at most 2-3% - see the Appendix). In another simplified intuitive picture, κ is scaled by how many more scattering events a phonon undergoes due to the crystalline boundaries within the length of its pristine material mean-free-path as:

$$\kappa = \frac{\kappa_0}{\left(1+\dfrac{\lambda_{pp}}{d}\right)} = \kappa_0 \frac{d}{\left[d+\lambda_{pp}\right]} \tag{15}$$

Note that $\Delta_{rms}$ or the boundary specularity does not appear in any of these models, which are assumed valid under diffusive phonon scattering conditions.

Figure 6a compares our Monte Carlo simulation results to those of the various nanocrystalline material models. We keep the temperature fixed at $T$ = 300 K and almost diffusive grain boundary scattering with $p$ = 0.1. With the exception of the model described by Eq. 12 which overestimates the thermal conductivity, and the model by Eq. 15 ("NC model") which underestimates it slightly at larger grain sizes, the models based on the simple reasoning of increased scattering rates and Kapitza resistance are in very good qualitative and quantitative agreement with the full Monte Carlo simulation results (blue line).

<u>Analytical models - ordered nanoporous case:</u> In the case of porous materials, the various analytical models are based on the simple logic that the thermal conductivity is reduced due to: i) the material volume reduction reflecting the reduction in the material heat capacity, and ii) the larger number of scattering events on the pore boundaries within the intrinsic phonon-phonon scattering mean-free-path length, similar to the nanocrystalline material case. A few commonly employed models in the literature for the thermal conductivity of nanoporous materials are given in the works of Eucken et al.,[75] Gesele et al.,[76] Tarkhanyan et al.,[80] Dettori et al. [Dettori15],[56] and Verdier et al.,[81] given by Eqs. 16 - 20 respectively:

$$\kappa = \kappa_0 \frac{(1-\phi)}{(1+\phi/2)} \quad (16)$$

$$\kappa = \kappa_0 (1-\phi) g_0^2 = \kappa_0 (1-\phi)^3 \quad (17)$$

$$\kappa = \kappa_0 \frac{(1-\phi)}{1+\left(\frac{\lambda_{pp}}{\delta}\right)} \quad (18)$$

$$\kappa = \kappa_0 \frac{(1-\phi)}{1+\frac{\phi}{2}+\frac{3\lambda_{PP}}{2D}} \quad (19)$$

$$\kappa = \frac{\kappa_0}{1+\frac{4}{3}\frac{\lambda_{PP}}{\delta}} \quad (20)$$

where $\delta$ is the average distance between adjacent pores and $D$ is the pore diameter. In Eq. 17 $g_0$ is related to percolation transport, approximated by the Looyenga effective medium model to be $(1 - \phi)^2$.[89] To extract the distance $\delta$, we determine the number of collision (scattering) events, $N_{coll}$, per unit length (along the length of the material towards the transport direction). The way that the number of collisions encountered is extracted, is simply by multiplying the size of the pores (area) by the number density of the pores ($\rho$) in domain in units of number/area, as:[56]

$$N_{coll} = \frac{\pi D^2}{4}\rho \quad (21)$$

The inverse of the number of interface scattering events per unit length provides the effective scattering distance $\delta$ between the pores ($\delta = 1/N_{coll}$). We adopted this from the works of Dettori et al.[56] and Lorenzi et al.[90] For instance, in the case of 10% porosity in the geometries we consider, the pores are spaced every 150 nm. The diameter is 50 nm, which from the above equation one can extract $\delta \sim 100$ nm, which is similar to an effective distance between the pore perimeters. In the case of 30% order porosity, for example this number changes to 39 nm.

Figure 6b shows a comparison of our diffusive boundary Monte Carlo simulations for the ordered porous structures of Fig. 3 "ordered cases", with the analytical models as described by Eqs. 16-20. The model of Gesele *et al.* (green line)[76] and Tarkhanyan *et* al. (purple line)[80] given by Eq. 17 and Eq. 18, respectively, show excellent agreement. The model by Dettori *et* al. (light-blue line)[56] given by Eq. 19 initially slightly underestimates the Monte Carlo results, but shows good agreement after $\phi = 20\%$. The model of Euken *et* al.[75] given by Eq. 16, which only accounts for the reduction in the material volume ($\phi$) in the first order, significantly overestimates the Monte Carlo results. Alternately, the model given by Eq. 20[81] (red line) accounts only for the mean-free-path reduction, but underestimates the reduction in $\kappa$ in the Monte Carlo results.

On the other hand, this very good match to ordered porous simulation data (especially of Eq. 17 and Eq. 18), signals that in the case of realistic variations and pore randomization, which have lower thermal conductivity, these models will fail, and more accurate models are needed. Indeed, the models described by Eq. 16 and Eq. 17 do not consider specific information regarding the details of the pore distribution in the material (positioning, diameters, shapes, interfaces, etc.), but only the volume reduction value. The models given by Eq. 18, Eq. 19, and Eq. 20 further consider an effective distance where scattering events are introduced, but still nothing about the distribution of those geometrical features in the material.

<u>Analytical models - random nanoporous case:</u> To construct an effective analytical model for the thermal conductivity in the case of structures with randomized pore geometries we consider the following logic: In the case of pores with randomized diameters and positioning, there are regions in the simulation domain that have a higher porosity than the average porosity of the overall material. These regions have an increased thermal resistance compared to the average resistance of the other segments of the material, something referred to as reduced 'line-of-sight'.[91,92,93] It is not clear though, how rearrangement of the thermal resistance along the length of the material in low and high resistance regions can affect the thermal conductivity and at what degree. In a previous work, for pores of constant diameter positioned randomly, we have shown that there is indeed a correlation between such rearrangements and lowering thermal conductivity.[31,91] On the other hand, by introducing a larger number of small diameter pores as in this work, the effect of resistance variation along the path is magnified since a reduced average diameter of pores provides greater surface area for phonon scattering.[32,60]

To construct an analytical model that can take this thermal resistance variation into account, we proceed as follows: We start by dividing the simulation domain into subdomains perpendicular to the transport direction, whose length $l_s$ is determined from scattering mean-free-path considerations using Eq. 21, i.e. $l_s = \delta$, as shown in Fig. 7a and 7b (for ordered and randomized geometries, respectively). Here, in the case of 10% porosity the $l_s \sim 100$ nm. We then compute the effective distance to adjoining pores $\delta_s$ separately for each subdomain of length $l_s$ again using Eq. 21. This is done over the total length of the simulation domain as shown in Fig. 7c and 7d, respectively. In the ordered case this remains at $l_s = \delta = 100$ nm (horizontal line in Fig. 7c). In the randomized pore geometry, however, $\delta_s$ deviates from $l_s$, as seen in Fig. 7d, especially in the regions of large deviation in local porosity. The local porosity profiles are also shown in Fig. 7e and Fig. 7f, respectively. In the ordered case the porosity averages to the global porosity (10%) every $l_s$, whereas in the disordered case the porosity deviates substantially, following the inverse trend of the distance deviation of Fig. 7d. For porosity we can construct higher resolution profiles as we have access to the porosity along the channel length limited by our domain discretization resolution. The red shaded regions in Fig. 7d and 7f depict the areas of distance/porosity smaller/greater than the average distance/porosity, which will introduce additional thermal resistance. We need to stress here that the choice of $l_s$, as the calculation of $\delta$ are extracted in a logical way, based completely on the underlying geometries. These change only when the porosity and randomness changes in the geometry, and are not parameters that we use arbitrarily to map the models to Monte Carlo data. Although the choice of $\delta$ as the distance between scattering events is intuitive, the choice of $l_s$ can also be justified by the fact that the important things that affect transport happen within the scattering lengths (see Appendix for sensitivity of the results in variations in $l_s$).

We then evaluate the standard deviation of the average scattering distance values along the length $\delta_s$, and label this as $\Delta\delta$. In the case of ordered pore arrangements this deviation would be zero. In the case of randomized pore geometries, however, it can be significant. To extract a more accurate value for $\Delta\delta$ we run 50 simulations of different randomized geometries for each porosity value and average the extracted 50 $\Delta\delta$. We then alter the distance $\delta$ in the models for random porosities as $\delta_r = \delta - \Delta\delta$ to account for an average smaller scattering mean-free-path. For example, in the case of 10% porosity with $\delta = 100$ nm, $\Delta\delta \sim 10$ nm, indicating an effective reduction in the distance between boundaries. In a similar way, the average deviation in porosity $\Delta\phi$ can be determined by using the porosity profiles along the length of the channel. We refer to these models from here on as the $\Delta\delta$ and the $\Delta\phi$ models.

As a first attempt to model thermal conductivity in randomized pore geometries we consider altering $\delta$ in the model by Tarkhanyan *et al.*[80] given by Eq. 18, which provides the best match to ordered porous simulations in Fig. 6b (purple line). The model now becomes:

$$\kappa = \kappa_0 \frac{(1-\phi)}{1+\left(\dfrac{\lambda_{pp}}{\delta-\Delta\delta}\right)} \quad (22)$$

In the simulated geometries, again the pore sizes are varied from $D = 10$ nm to 50 nm in a random fashion using a uniform distribution. The thermal conductivity predictions are plotted by the black line in Fig. 8, which compares this model with the full Monte Carlo results (blue line). For reference, we plot by the purple-dashed line the result of the same model in the ordered case as in Fig. 6b, which significantly overestimates the thermal conductivity. As we can see, the improved model has very good agreement with the MC results for high porosities, but for low porosities still some mismatch is observed.

In order to improve the $\Delta\delta$ model (Eq. 22) for low porosities, we consider further the effect of local resistance imposed by the geometrical arrangement. Clearly, a large number of pores would give a very small local $\delta_s$, i.e. high local porosity and will impose a significant degradation of thermal conductance. Thus, the model is extended to include the possibility that some subdomains can have a porosity much higher than the average porosity, i.e. regions of high thermal resistance contributing to a more substantial drop in $\kappa$. For this, we simply consider the deviation in the scattering distances in each subdomain as before, but we now weigh more the regions where the local $\delta_s$ is less than that of the uniform case. In this way, we increase the dominance of regions with higher local porosity in determining thermal resistance. Thus, the deviation in the scattering lengths/porosity, decreases/increases even further as:

$$\Delta\delta_w = \Delta\delta \frac{L_X}{L_X - L_H} \tag{23}$$

where $L_H$ is the proportion of these high porosity regions compared to the overall length of the domain $L_X$. For example, for porosities $\phi$ = 10%, 20%, 30%, $\Delta\delta_w$ ~ 19 nm, 10 nm, and 8 nm, respectively, an increase compared to the corresponding non-weighted $\Delta\delta$ ~ 10 nm, 8 nm, and 8 nm. Thus, an improved model (will refer to it as '$\Delta\delta_w$ model') is now given by:

$$\kappa = \kappa_0 \frac{(1-\phi)}{1 + \left(\dfrac{\lambda_{pp}}{\delta - \Delta\delta_w}\right)} \tag{24}$$

As can be seen in Fig. 8, this model (red line) has very close agreement with the full Monte Carlo results (blue line) for the random porosity case. We note that considering both the weighted deviation in distance and porosity as:

$$\kappa = \kappa_0 \frac{(1-(\phi + \Delta\phi_w))}{1 + \left(\dfrac{\lambda_{pp}}{\delta - \Delta\delta_w}\right)} \tag{25}$$

seems to underestimate the thermal conductivity predictions compared to Monte Carlo (green line in Fig. 8), possibly because of double counting the scattering events. We note that the choice of the regions $L_H$ that have in absolute terms smaller $\delta_s$ (or higher porosity) than the average $\delta$, is arbitrary. One can consider $L_H$ as regions which have $\delta_s$ smaller than a fraction of $\delta$, which might be more relevant when the variations are very small. In our case, however, the variation is substantial and the reference we used ($\delta$) provided a good match with the Monte Carlo simulation results.

Next, as a second example, we test the same methodology by increasing the effective porosity in the model by Gesele *et al.*[76] in Eq. 17. We increase the porosity by the overall deviation $\Delta\phi$, and by the weighted deviation $\Delta\phi_w$ (referring to them as the '$\Delta\phi$' and '$\Delta\phi_w$' models). Here we only modify the transport component in the model, i.e.:

$$\kappa = \kappa_0 (1-\phi)(1-(\phi+\Delta\phi))^2 \qquad (26)$$

$$\kappa = \kappa_0 (1-\phi)(1-(\phi+\Delta\phi_W))^2 \qquad (27)$$

Figure 9 shows the comparison of these models to our Monte Carlo results (blue line) versus porosity. Since we simulated 50 different channels for each porosity value, in the model we use the average $\Delta\phi_w$ value for all 50 of these geometries. In the case of 10%` porosity, for example (shown in Fig. 7a), $\Delta\phi$ increases the effective porosity by another 7.5% to the total 17.5% porosity. The model predictions using $\Delta\phi$, given by Eq. 26, are shown by the black line in Fig. 9. While this model works well for higher porosity values above $\phi = 20\%$, it again overestimates the thermal conductivity for porosities below $\phi = 20\%$. This is again due to the possibility of regions of porosities above the average value in the simulation domain, which dominate thermal resistance. At higher porosities the pores make the different regions look more uniform. However, this is significant improvement compared to the starting model of Gesele *et* al. Eq. 17, which as shown by the green-dashed line, it significantly overestimates the thermal conductivity. Note that in the uniform porosity cases $\Delta\phi$ is effectively zero.

In the case where we include the weighted porosity $\Delta\phi_w$ (the '$\Delta\phi_w$ model' given by Eq. 27), the model is significantly improved as indicated by the red line in Fig. 9, which essentially overlaps the Monte Carlo results (blue line). Here, for the 10% porosity case, $\Delta\phi_w$ is computed to be 18% after we average it over 50 different channels. This is a ~10% increase in the overall porosity over the non-scaled $\Delta\phi$ above. This makes the effective porosity of this disordered channel to increase by almost a factor of three to $\phi+\Delta\phi_W = 28\%$. In summary, it turns out that for $\phi = 10\%$, 20%, and 30%, $\Delta\phi_w = 18\%$, 21%, and 16%. As porosity increases the deviation in porosity, in general, decreases. However, these numbers stress the importance of nanostructured geometry variability in thermal conductivity, an effect that is usually overlooked, but increases the effective porosity significantly, with its effect being more dominant even than boundary roughness.

Thus, the new models ($\Delta\delta_w$ and $\Delta\phi_w$) described by Eq. 24 and Eq. 27 are shown to be accurate and can be used to extract first order thermal conductivity estimations for structures with random pore positions and diameters, using knowledge of basic geometrical specifics in the nanostructure. In an experimental setting, these models can not only be used to understand thermal conductivity measurements if the geometrical features are known, but also conversely, to estimate the degree of disorder in nanostructures, once average porosity and thermal measurement data is available. Details on the degree of randomization might be hard to extract in the entire domain of the material, but these models provide a possible way of estimating this. Finally, we note that recent theoretical studies about the effect of the effect of variations for electronic transport (where the mean-free-path is much shorted), is not as noticeable.[6,94] Thus, variability can be used as means to achieve lower thermal conductivity without affecting the electronic system, which could be advantageous for thermoelectric applications.

<u>Verifying Matthiessen's rule</u>: When different scattering events are considered independently, it is usual practice to combine the different scattering rates, or resistivities using Matthiessen's rule. It is important to examine if the combination of nanocrystallinity and nanoporosity can be combined using Matthiessen's rule, which will provide an indication of the degree of independence of the two scattering mechanisms. Here, we examine this using the Monte Carlo simulation results for each mechanism independently and together. We also examine the eligibility of the analytical models we have constructed in being used within the Matthiessen's rule. In this case we consider scattering due to: i) three phonon processes leading to a $\kappa_{PH}$, ii) scattering due to nanocrystalline geometries leading to a $\kappa_{NC}$, and iii) scattering due to nanopores leading to a $\kappa_{NP}$. Thus, the total conductivity is given by:

$$\frac{1}{\kappa_{TOTAL}} = \frac{1}{\kappa_{PH}} + \frac{1}{\kappa_{NC}} + \frac{1}{\kappa_{NP}} \qquad (28)$$

We proceed with our verifications as follows: i) We simulated structures including nanocrystallinity with $<d>$ = 225 nm and nanoporosities 5%, 10%, and 15%. In the ordered pore case we use $D = 50$ nm and in the random pore case $D$ varies from $D = 10$ nm to 50 nm in a random fashion using a uniform distribution. ii) We simulated the nanocrystalline geometry structures and the nanoporous structures separately, for the same $<d>$ and $\phi$ as in (i). iii) We computed the combined thermal conductivity of the two simulations in (ii) using Matthiessen's rule and compared that to the combined Monte Carlo simulation of (i). iv) We combined the results of the analytical models for nanocrystallinity (the "NC model" given by Eq. 15) and porosity ($\Delta\delta_w$ and $\Delta\phi_w$ models given by Eq. 24 and Eq. 27 respectively) and compared those to the Monte Carlo simulation results of (i). For the structures with porosities $\phi$ = 5 %, 10 %, 15 %, we extracted $\Delta\delta_w$ = 27 nm, 19 nm, 13 nm, and $\Delta\phi_w$ = 20%, 18%, 16%, respectively.

The summary of these comparisons is shown in Fig. 10a and 10b for the ordered and randomized porous materials, respectively. The blue bars show the full MC simulations which include nanocrystallinity and porosity. The error bars indicate the spread of the 50 simulations performed to extract the average value of the thermal conductivity. The green bars indicate MC simulations for each scattering environment separately (nanocrystallinity and porosity), coupled together using Matthiessen's rule to extract the overall thermal conductivity. The purple bars indicate the thermal conductivity predicted by the $\Delta\delta_w$ analytical model (Eq. 24) combined with the nanocrystalline model by Eq. 15 using Matthiessen's rule. Finally, the red bars indicate the thermal conductivity predicted by the combination of the $\Delta\phi_w$ model (Eq. 27) and Eq. 15 using Matthiessen's rule. The percentage difference of the three latter situations compared to the full MC results is indicated on top of the respective bars. Clearly, a very good agreement between the full MC results, the partial MC results, and the analytical models is observed. In the ordered pore case shown in Fig. 10a the error introduced by the analytical models is at most 10%, observed for the 15% porosity material. In the case of random pores results shown in Fig. 10b, a slightly larger variation is observed for the larger porosities, but it is still at most 13 % (for both partial MC results and the analytical models).

The good agreement between all results, indicates that Matthiessen's rule is still valid at large degree and the nanocrystallinity and nanoporosity can be treated as independent mechanisms. It also indicates that well-validated analytical models are at first order accurate for estimating phonon transport in complex nanostructured materials, at least silicon at room temperature. Quantifying the validity of Matthiessen's rule is especially important for experimentalists seeking a fast verification of measured data, but we should note that our conclusions could be only confidently valid for the structures and geometries we considered. Deviations from Matthiessen's rule have been observed in various cases for phononic, but also electronic systems.[95,96] In particular, we only considered nanopores larger than 10 nm in diameter to stay within the validity of the particle nature of phonons as treated by Monte Carlo. Other works, however, have considered smaller pores ($D$ less than 10 nm), which also drastically reduce $\kappa$, but indicate larger violations of Matthiessen's rule.[56,81,90,97,98,99]

## V. Conclusions

In this work we have developed and employed a 'single-phonon' Monte Carlo phonon transport simulator to solve the Boltzmann Transport Equation for phonons in hierarchically disordered Si nanostructures. We investigated the presence of nanocrystalline and nanoporous features separately and combined, in ordered and disordered realizations. In nanocrystalline geometries the effect of grain size on $\kappa$ is more pronounced at grain sizes <$d$> smaller than the average phonon mean free path of the system ($\lambda_{pp}$). In that case, boundary scattering dominates over internal three-phonon scattering. We further show that the effect of changing porosity ($\phi$) on thermal conductivity is much larger than boundary roughness and specularity ($p$) in reducing $\kappa$. An important result of this work is that it demonstrates that randomization in disorder, which is often overlooked, can play an important effect, further reducing thermal conduction by even up to 60% compared to the ordered pore geometry. Thus, non-uniformity can be as important, if not more important in reducing $\kappa$ compared to boundary roughness and specularity ($p$) and needs to be considered at a similar level in interpreting experimental data. Based on simple geometrical rules and previous analytical models for ordered structures, we constructed accurate analytical models for randomized porous structures with excellent agreement with the full scale Monte Carlo simulations. We believe our results and the models presented will provide guidance in developing better understanding of thermal transport in nanostructured materials and aid the design of better thermoelectric and heat management materials.

## VI. Acknowledgements

This work has received funding from the European Research Council (ERC) under the European Union's Horizon 2020 Research and Innovation Programme (Grant Agreement No. 678763).

[42]  K. Termentzidis, M. Verdier, and D. Lacroix, Effect of Amorphisation on the Thermal Properties of Nanostructured Membranes, Zeitschrift für Naturforschung A 72.2, 189, (2017).

[43]  D. Vasileska, K. Raleva, S. M. Goodnick, Z. Aksamija and I. Knezevic, Thermal modeling of nanodevices, 14th International Workshop on Computational Electronics, Pisa (IEEE, Pisa, 2010), Vol. 355, pp 1-5.

[44]  E. Pop, S. Sinha, and K. E. Goodson, Thermal Phenomena in Nanoscale Transistors, J. Electron. Packag. 128, 102 (2006).

[45]  S. Mazumdar, and A. Majumdar, Monte Carlo Study of Phonon Transport in Solid Thin Films Including Dispersion and Polarization, J. Heat Transfer, 123, 749, (2001).

[46]  E. Pop, R. W. Dutton, and K. E. Goodson, Analytic band Monte Carlo model for electron transport in Si including acoustic and optical phonon dispersion, J. Appl. Phys., 96, 4998, (2004).

[47]  Q. Hao, G. Chen, and M.S. Jeng, Frequency-dependent Monte Carlo simulations of phonon transport in two-dimensional porous silicon with aligned pores, J. Appl. Phys. 106, (2009).

[48]  S. Wolf, N. Neophytou, Z. Stanojevic, and H. Kosina, , Monte Carlo Simulations of Thermal Conductivity in Nanoporous Si Membranes, J. Electron. Mater., 43, 3870, (2014).

[49]  A. Mittal and S. Mazumder, Monte Carlo Study of Phonon Heat Conduction in Silicon Thin Films Including Contributions of Optical Phonons, Journal of Heat Transfer 132, 052402 (2010).

[50]  S. V. J. Narumanchi, J. Y. Murthy, and C. H. Amon, Comparison of Different Phonon Transport Models for Predicting Heat Conduction, J. Heat Transfer 127, 713 (2005).

[51]  C. Jeong, S. Datta and M. Lundstrom, Monte Carlo Study of Phonon Heat Conduction in Silicon Thin Films Including Contributions of Optical Phonons,  J. Appl. Phys., 111, 093708, 2012).

[52]  D. P. Sellan and E. S. Landry and J. E. Turney, A. J. H. McGaughey and C. H. Amon. Size effects in molecular dynamics thermal conductivity predictions, Phys. Rev. B. 81, 214305 (2010).

[53]  L. Weber, E. Gmelin, Transport Properties of Silicon, Appl. Phys. A 53,136-140

(1991).

[54]  Y. S. Ju, and K. E. Goodson, Phonon scattering in silicon films with thickness of order 100 nm, Appl. Phys. Lett., 74(20), 3005 (1999).

[55]  G. Chen, Thermal Conductivity and Ballistic Phonon Transport in Cross-

PlaneDirection of Superlattices, Phys. Rev. B 57, 14958 (1998).

[56]  R. Dettori, C. Melis, X. Cartoixà, R. Rurali and L. Colombo, Model for thermal conductivity in nanoporous silicon from atomistic simulations, Phys. Rev. B- Condens. Matt. Mater. Phys., 91, 054305, (2015).
29

Figure 1:

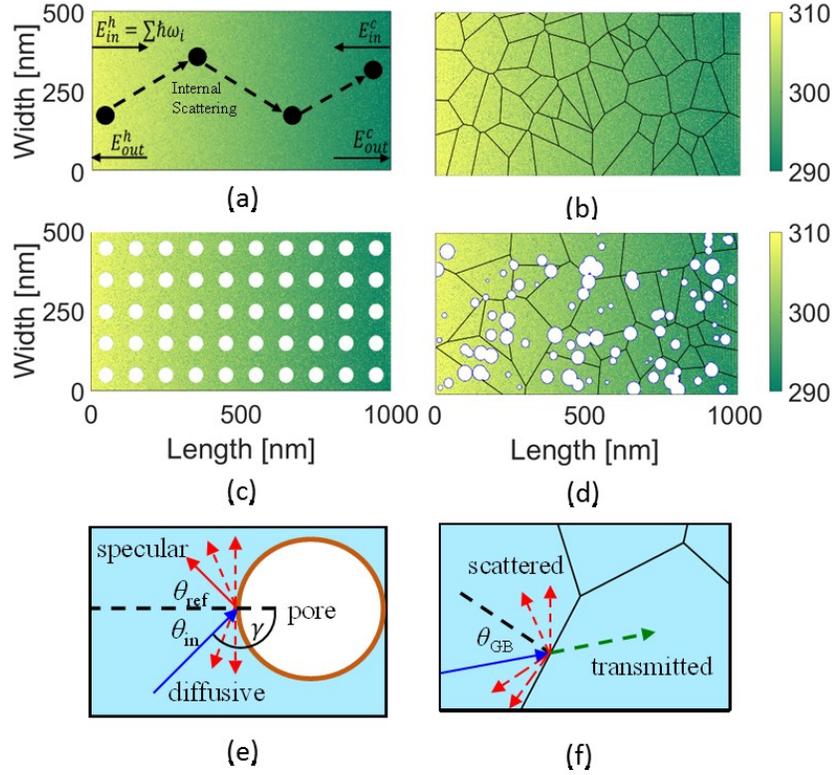

Figure 1 caption:

Examples of the nanostructured geometries considered. The coloring indicates the established thermal gradients when the left and right contacts are set to $T_H$ = 310 K **(yellow)** and $T_C$ = 290 K **(green)**. (a) Pristine silicon channel. (b) Nanocrystalline (NC) channel. (c) Ordered nanopores (NP) within the channel material of ~20% porosity in a rectangular arrangement. (d) Combined NC and disordered NP material. (e) Schematic of scattering mechanism for pore scattering, indicating the pore boundary, the initial angle of the phonon $\theta_{in}$, and potential new angle of propagation $\theta_{ref}$ depending on specularity parameter $p$. Probable paths of the phonon after scattering for both diffusive (red dashed lines) and specular (red solid line) are depicted. (f) Schematic of the scattering mechanism for grain boundary scattering, indicating the initial angle of the phonon $\theta_{GB}$ from the normal (dashed line), grain boundaries (black lines), initial path of the phonon (blue line) and probable paths of the phonon after scattering [red dashed lines and green dashed (transmitted) line].

Figure 2:

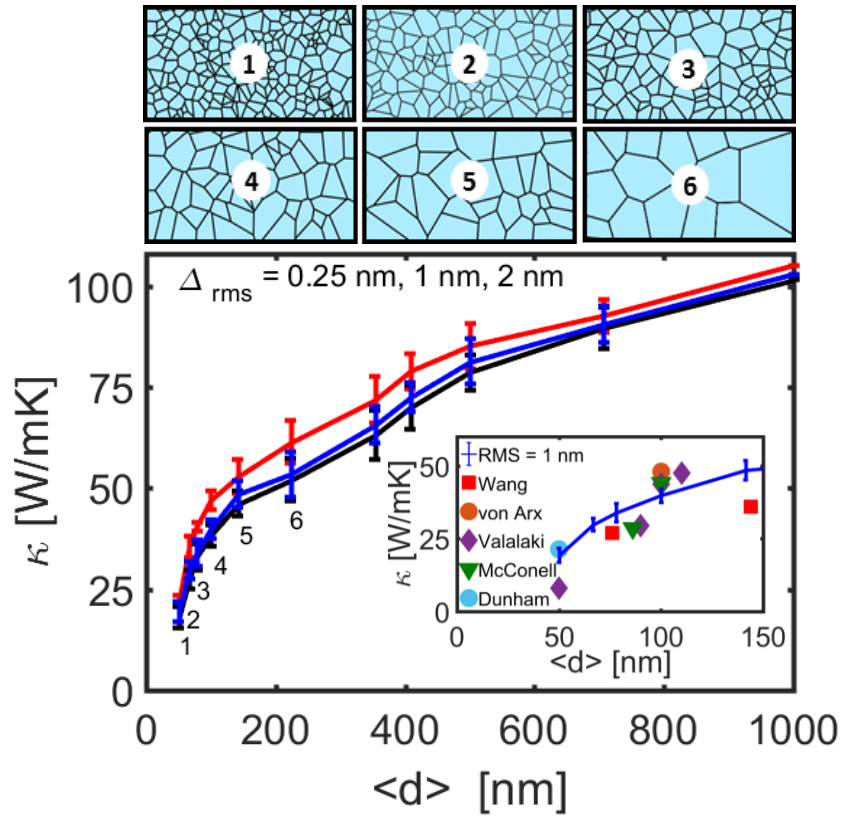

Figure 2 caption:

The effects of grain size and grain boundary roughness ($\Delta_{rms}$) on the thermal conductivity of the silicon channel. Grain size is varied from an average grain dimension <d> of 1000 nm down to 50 nm. The structure geometry insets labelled "1" to "6" give typical examples of geometries from <d> = 50 nm to 225 nm, respectively. We simulate three different values of grain boundary roughness, $\Delta_{rms}$ = 0.25 nm (red line), 1 nm (blue line) and 2 nm (black line). Each point is an average of 50 simulations. A sharp drop in thermal conductivity is observed below <d> ~ 140 nm (structure sub-figure and point "4"). Inset: Some available experimental results[14,21,68,73,74] are compared to the $\Delta_{rms}$ = 1 nm (blue line).

Figure 3:

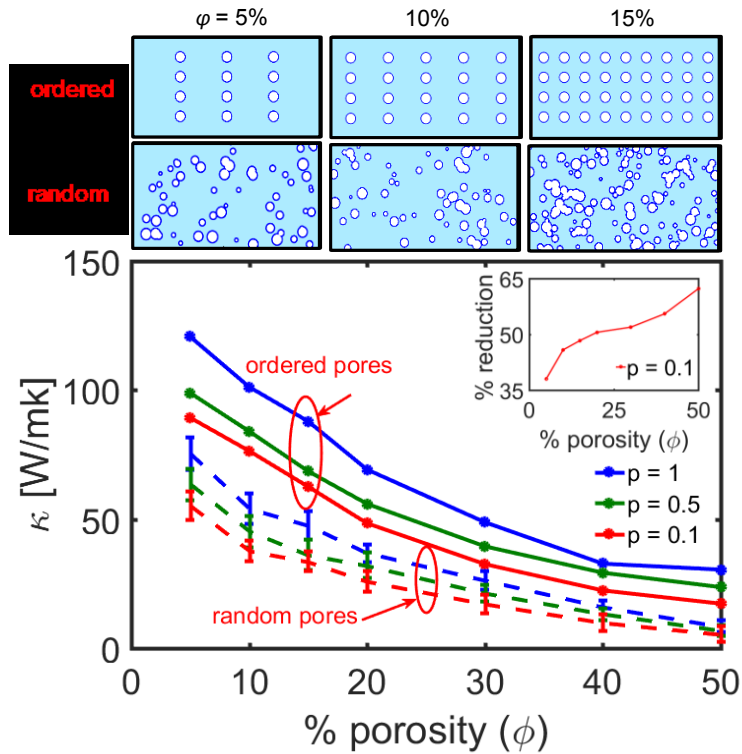

## Figure 3 caption:

The thermal conductivity versus porosity ($\phi$) for two geometry cases - ordered case (solid lines) and random case (dashed lines). Three different values for boundary specularity are considered: $p = 1$, totally specular boundary scattering (blue lines); $p = 0.5$ (green lines); and $p = 0.1$, almost diffusive boundary scattering (red lines). The inset depicts the percentage reduction in thermal conductivity for the $p = 0.1$ (red line), random porosity case compared to the ordered case. The geometry structures of the simulated geometries for ordered and random arrangement cases for 5%, 10 % and 15 % porosity are shown on top of the figure. In all cases the domain size is fixed to length $L_x = 1000$ nm and width $L_y = 500$ nm.

Figure 4:

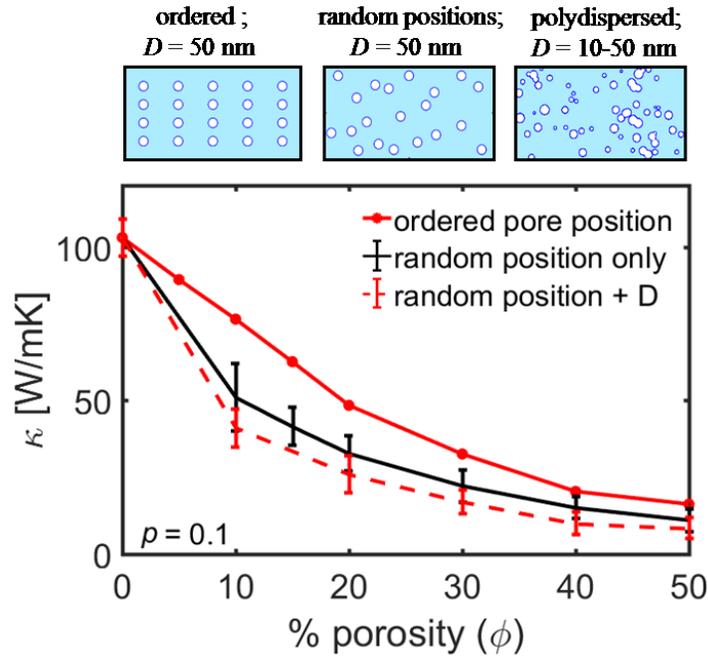

Figure 4 caption:

The thermal conductivity versus porosity ($\phi$) for ordered pore structures, randomized pore structures, and polydispersed geometries with randomized pore positions and diameters. In the first two cases (red and black solid lines), the diameter is fixed at $D$ = 50 nm (see schematics panel for 10% porosity). In all cases the specularity for all boundaries is fixed at $p$ = 0.1

Figure 5:

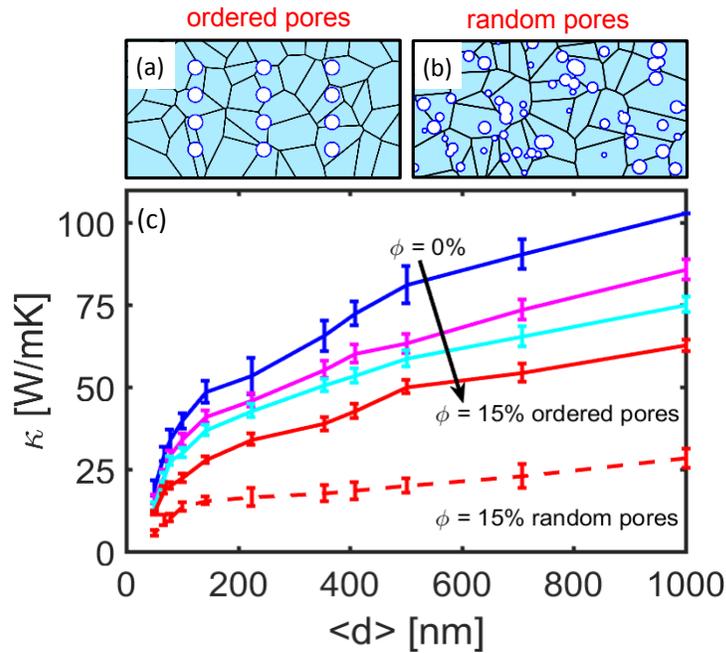

## Figure 5 caption

Monte Carlo simulations showing the combined effects of grain size and porosity ($\phi$) in both the ordered pores case (solid lines) and random pores case (dashed line) versus grain size <d>. The thermal conductivity at for porosities $\phi$ = 0%, 5%, 10% and 15% are shown by the blue, magenta, light blue, and red lines respectively. The effect of combined nanocrystalline and nanoporous material with random pore positions and sizes (uniformly distributed between 10 nm to 50 nm) is depicted by the red-dashed line. Examples of typical geometries simulated for the case of 5% porosity, for both ordered and random pore arrangements, are shown above the figure.

Figure 6:

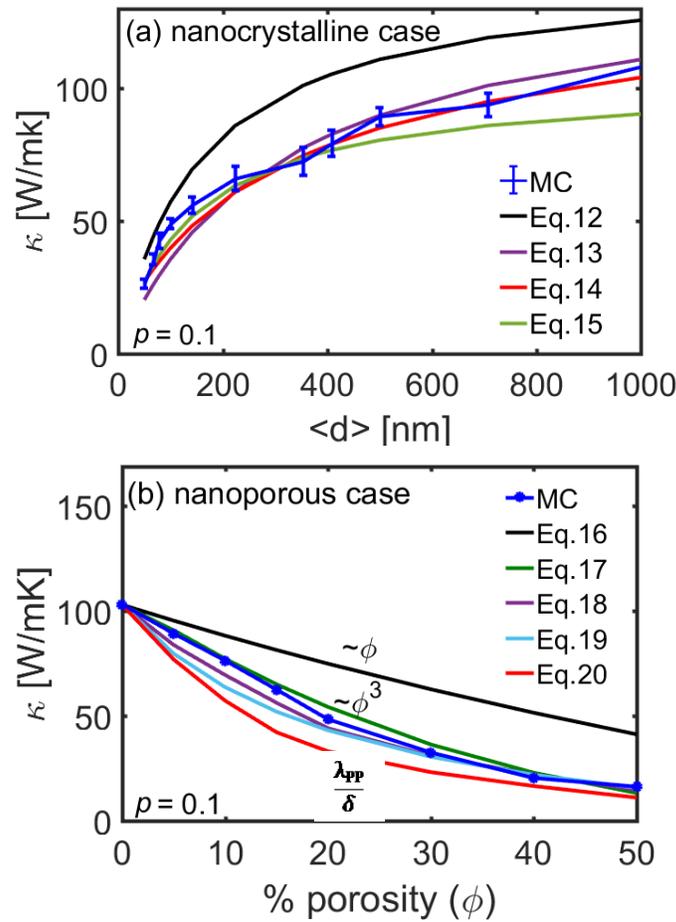

Figure 6 caption:

(a) Thermal conductivity versus grain size for commonly employed analytical models for nanocrystalline geometries compared to the Monte Carlo results of this work (blue line). The grain size is varied from an average of $<d>$ = 1000 nm down to 50 nm with a roughness $\Delta_{rms}$ = 1 nm. (b) Thermal conductivity versus porosity for the commonly employed analytical porous material models compared to the Monte Carlo results of this work (blue line). The pore boundary specularity is fixed at $p$ = 0.1. In both cases the domain top/bottom roughness specularity is set to $p$ = 0.1.

Figure 7:

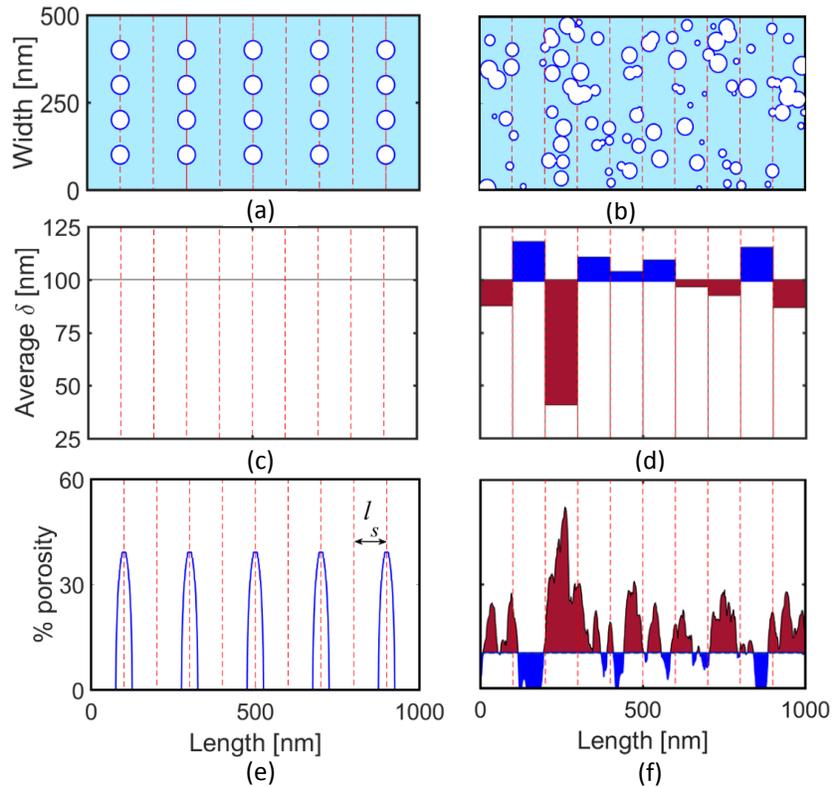

Figure 7 caption:

Extraction of the variation in distances between pores, $\delta$, and variation in porosity, $\phi$, in the nanoporous materials examined. Geometries for ordered (a) and randomized (b) nanoporous geometries with $\phi$ = 10 % are shown on top. The distribution of distances between pores, averaged every $l_s$ = 100 nm (depicted by the dotted red lines), is shown in (c) and (d), respectively. The distribution of pore distances is well defined and constant in the ordered case, but deviates in the randomized pore geometry. The distribution of porosity is shown in (e) and (f), respectively. In this case the distribution can be evaluated with higher resolution along the length of the material. In (e) the porosity averages to $\phi$ = 10 % in every $l_s$ = 100 nm domain. In (f) the porosity deviates from the 10 % average following an inverse trend compared to the distance between the pores shown in (d). The red shaded portions of the distance profile in (d) and the porosity profile in (f) represent the regions of increased thermal resistance.

Figure 8:

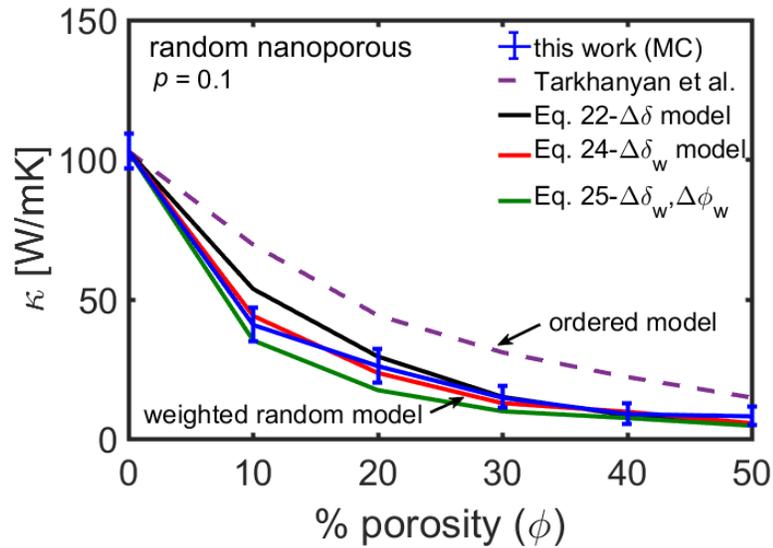

Figure 8 caption:

Thermal conductivity versus porosity for the analytical models of randomized pore geometries, compared to the Monte Carlo simulation results of this work (blue line). Pore boundary specularity in Monte Carlo is fixed at $p = 0.1$. The model of Tarkhanyan *et al.*[80] as described by Eq. 18 is shown by the dashed-purple line. Equation 22 (black line) incorporates a deviation $\Delta\delta$ in the average distance. There is good agreement with Monte Carlo results for porosities beyond $\phi = 20\%$. To improve the model, Eq. 24 (red line) incorporates a weight on the deviation $\Delta\delta_w$, increasing the importance of regions of higher porosity. As a reference, Eq. 25 (green line), incorporates a further weighted deviation in porosity $\Delta\phi_w$, which, however, slightly underestimates the Monte Carlo simulations.

Figure 9:

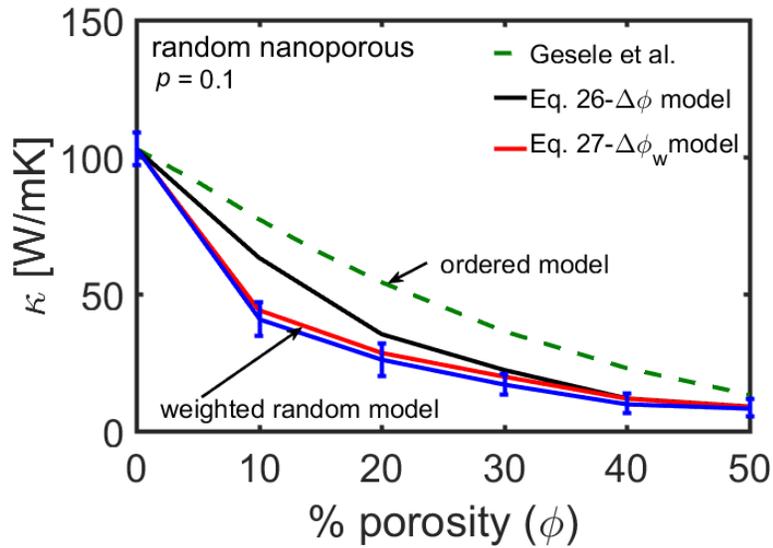

Figure 9 caption:

Thermal conductivity versus porosity for the analytical models of randomized pore geometries compared to the Monte Carlo simulation results in this work (blue line). Pore boundary specularity in Monte Carlo is fixed at $p = 0.1$. The model of Gesele et al.[76] as described by Eq. 17 is shown by the dashed-green line. Equation 26 (black line) incorporates a deviation $\Delta\phi$ in the average porosity. There is good agreement with the Monte Carlo results for porosities beyond $\phi = 30\%$. To improve the model, we incorporate a weight on the porosity $\Delta\phi_w$ (Eq. 27) increasing the importance of regions of higher porosity (red line).

Figure 10

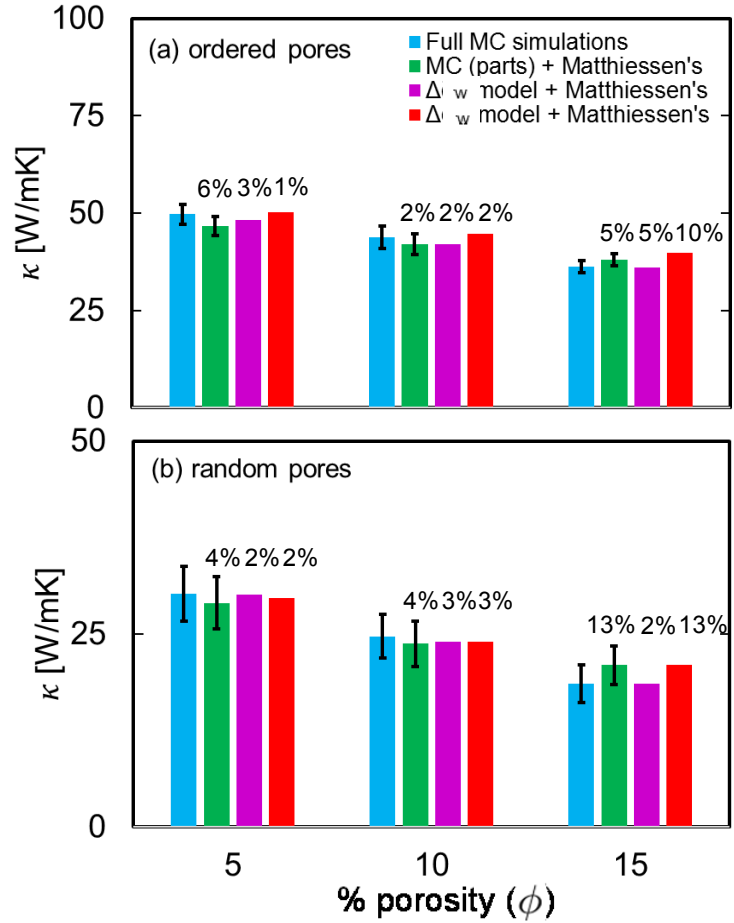

Figure 10 caption:

Comparison between the full Monte Carlo (MC) simulated results in structures with grains and pores (blue bars) and: i) MC simulation results of grains alone and pores alone, but combined through Matthiessen's rule (green bars), ii) results given by the porous material model introduced in Eq. 24 ($\Delta\delta_w$ model) combined with the nanocrystalline model of Eq. 15 through Matthiessen's rule (purple bars), and iv) results given by model introduced in Eq. 27, ($\Delta\phi_w$ model) combined with the nanocrystalline model of Eq. 15 through Matthiessen's rule (red bars). (a) Ordered pore geometries. For the MC simulations, 50 realizations with grain boundaries of $<d>$ = 225 nm are averaged, and pores of a fixed diameter $D$ = 50 nm. (b) Randomized pore geometries. The pore diameters vary from 10 nm to 50 nm. The percentage numbers indicate the variation of each method from the full MC results (blue bars). Porosities $\phi$ = 5%, 10 % and 15% are shown.

# Appendix:

In this Appendix we include sensitivity studies to the simulation parameters and model choices we employ in the paper. The purpose is to demonstrate that the conclusions we reach in the paper, and the analytical models we propose, are qualitatively and quantitatively robust with respect to model assumptions and parameter choices. The following parameters are examined:

## 1. Phonon-phonon mean-free-path (mfp) value:

The influence of a different choice for the phonon-phonon scattering mean-free-path $\lambda_{pp}$ when scaling the simulated thermalconductivity in Eq. 8. In the literature $\lambda_{pp}$ varies from 100 nm to 300 nm, thus, here, we recreate Fig. 5 of the main text as Fig. A1 for $\lambda_{pp}$ = 135 nm – solid lines (as in the main text) and $\lambda_{pp}$ = 300 nm – dashed lines are used in Eq. 8. Figure A1 shows the thermal conductivity versus nanocrystalline domain size for porous materials with porosities $\phi$ = 0% and 15%, in both ordered and randomized pore conditions. Doubling the mfp has at most ~15% qualitative difference in our results in the pristine material with no crystallinity and no porosity (compare the dashed to solid blue lines at $<d>$ = 1000 nm), which drops to ~6% in the case where high and randomized disorder is introduced (dashed versus solid purple lines at $<d>$ = 1000 nm). At smaller $<d>$ the dependence on mfp is insignificant, indicating that boundary scattering dominates transport. Thus, the assumption of mfp choice in Eq. 8 does not change any of our quantitative or qualitative trends.

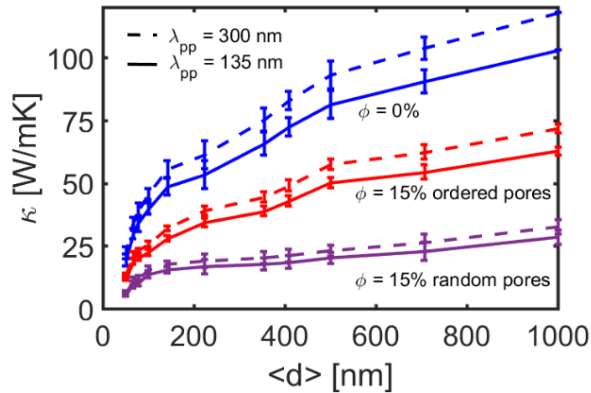

**Figure A1:** Monte Carlo simulations showing the effect of different phonon-phonon scattering mean-free-paths on the combined effects of grain size $<d>$ and porosity ($\varphi$). The blue lines show the thermal conductivity in the presence of nanocrystallinity only (no pores). Ordered pores case (red lines) and random pores case (purple lines) versus grain

size <d> are also shown. Porosity $\varphi = 15\%$ is considered. The dashed lines indicate the simulations where $\lambda_{pp} = 300$ nm.[44,53] The solid lines are for $\lambda_{pp} = 135$ nm as in the main text.[51,57]

## 2. Constant roughness $\Delta_{rms}$ versus constant specularity $p$:

Instead of a constant specularity $p$, in Monte Carlo it is also customary to determine the actual specularity for each phonon using the expression $p(q) = \exp(-4q^2\Delta_{rms}^2)$, which also allows wavevector dependence reflections. In that case, what is constant is the surface roughness ($\Delta_{rms}$). Below, we recreate Fig. 3 of the manuscript (for the ordered pore cases only) as Fig. A2, but include simulation results for constant $\Delta_{rms} = 0.3$ nm treatment of pore boundary scattering (i.e. now there is q-dependent scattering), a value which corresponds well to rough silicon surfaces.[65,66] This specific $p = 0.1$ (red line in Fig. A2) we employ throughout the main text, seems to correspond to this $\Delta_{rms} \sim 0.3$ nm in all the porosity values we consider (black-dashed line in Fig. A2). This means that the average phonon wavevector (from the expression above) can be extracted to be $q = 2.5$/nm, which corresponds to phonons around the first quarter of the Brillouin zone (length $2\pi/a_0$, where $a_0 = 0.543$ nm).

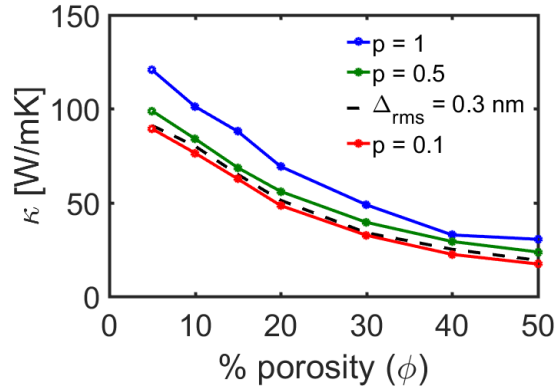

**Figure A2:** Comparison of fixed specularity values vs fixed $\Delta_{rms}$ (black line) for ordered porous geometry cases. Three different values for fixed boundary specularity are considered: $p = 1$, totally specular boundary scattering (blue line); $p = 0.5$ (green line); and $p = 0.1$, almost diffusive boundary scattering (red line). The results for the fixed $\Delta_{rms} = 0.3$ nm (black-dashed line) most closely correspond to $p = 0.1$.

## 3. Channel length dependence:

Throughout the paper, we have fixed the channel length at $L_x = 1000$ nm, which is indeed shorter than some of the phonon mean-free-paths in Si, and used a scaling to adjust for this short channel described by Eq. 8 in the main text. Here we performed Monte Carlo simulations in nanoporous materials of channel length twice as much, at $L_x = 2000$ nm and compare the thermal conductivity results for the two cases. Figure A3 shows the comparison of the thermal conductivity versus porosity in channels with the different lengths, $L_x = 1000$ nm (blue line), and $L_x = 2000$ nm (red line). Indeed, due to the scaling performed, the boundary scattering on the upper/lower surfaces, as well as scattering on the pores, the channel we consider ($L_x = 1000$ nm) is already diffusive, and changing the length does not alter the thermal conductivity, either for the pristine channel (for $\phi = 0$) or for the porous channels at any porosity.

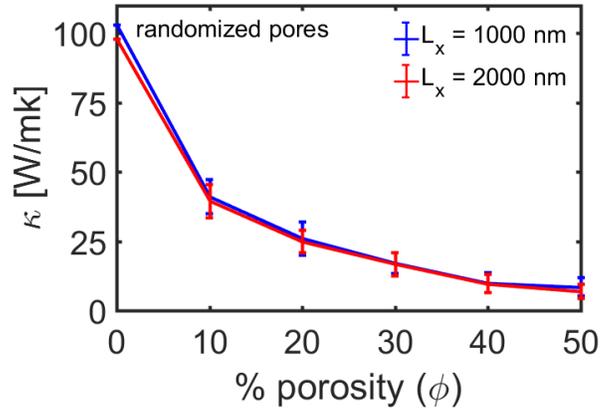

**Figure A3:** (b) Thermal conductivity versus porosity for randomized nanoporous geometries in the nominal domain length $L_x = 1000$ nm (blue line) and doubled domain length $L_x = 2000$ nm (black line). Pore boundary specularity is $p = 0.1$.

**4. Grain size and average phonon path:**

In order to have a clear picture of the transport regime at which the channel operates (ballistic versus diffusive), we have also calculated the average phonon path length in our polycrystalline structures from the moment the phonons enter the domain after initialization, to the moment they are extracted from the domain. This is shown below in Fig. A4 versus the nanocrystalline size $<d>$. The average phonon path length in the pristine channel is only twice the length of the domain (at $<d> = 1000$ nm), which is the reason for the phonon mean-free-path scaling we employ by Eq. 8 As the nanocrystallites are reduced in size, the path increases, especially when their size becomes smaller compared to $\lambda_{pp}$ (135 nm). The path of the phonons is then more than an order of magnitude compared to the channel length, indicating compete channel diffusion, and large reductions in the thermal conductivity.

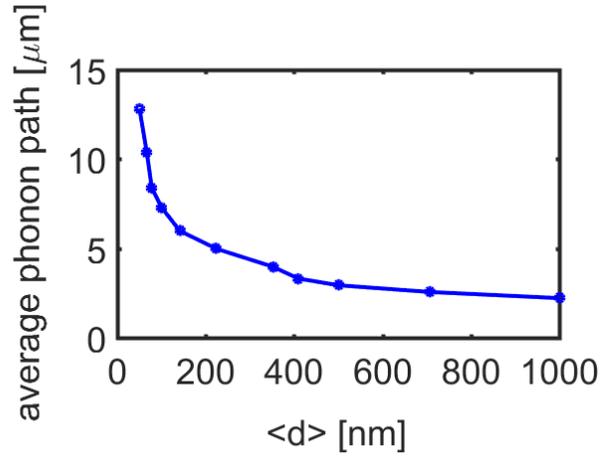

**Figure A4: (a)** Average phonon path vs <d> for nanocrystalline geometry cases. <d> is varied from <d>=1000 nm to <d> = 50 nm.

## 5. Kapitza resistance variation:

In the analytical models for nanocrystalline materials described by Eqs. 12, 13, 14, the value of the Kapitsa resistance appears. There is a slight variation in the values of the Kapitsa resistance in the literature, from $R_K$ = 1-1.16 $10^9$ Km$^2$W$^{-1}$. Here we vary the value of $R_K$ in that range to examine the amplitude of this variation in the thermal conductivity. Indeed, the effect of this variation, as shown in Fig. A5 is minor, both qualitatively and quantitatively.

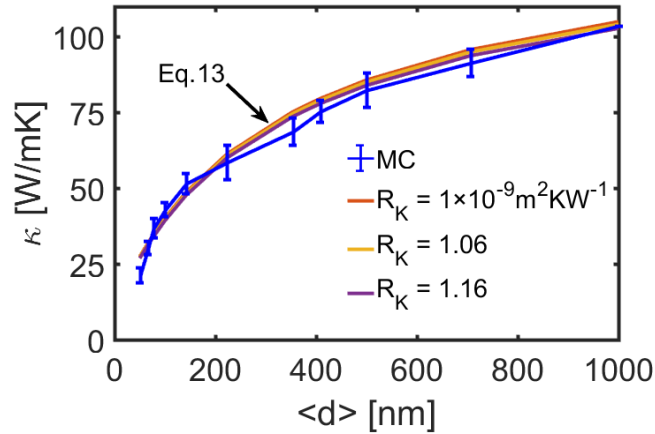

**Figure A5:** Thermal conductivity versus grain size from the commonly employed Yang model[79] analytical model for nanocrystalline geometries compared to the Monte Carlo results of this work (blue line). We assume $\Delta_{rms}$ = 1 nm. The Kapitza resistance value is varied from 1 x $10^9$ Km$^2$W$^{-1}$ (red line), to 1.06 x $10^9$ Km$^2$W$^{-1}$ (yellow line) to 1.16 x $10^9$ Km$^2$W$^{-1}$ (purple line).

## 6. Choice of domain splitting $l_s$ for the randomized analytical models:

In the extension of the analytical models in order to capture the effect of randomized porosity (Eqs. 24, 27), we split the simulation domain in lengths of $l_s = \delta$, where $\delta$ is the scattering length introduced by the pores, and determine the deviation in porosity across the length of the channel based on that $l_s$ region separation. Although $\delta$ is determined solemnly by the underlying geometry, $l_s$ is a choice we make based on the fact that the effect of porosity will be correlated to the scattering distance it causes. However, here we investigate the sensitivity of the proposed models on the choice of $l_s$. We separated the domain in $l_s = \delta$, $l_s = 2\delta$ and $l_s = \delta/2$, and extracted the deviations in porosity based on those separation. We then included them in the analytical model given by Eq. 24. Figure A6 below is a recreation of Fig. 8 of the main text, which shows that: i) independent of the choice of $l_s$, the model that included deviations provides better fit to the Monte Carlo data compared to the simple, non-randomized model (dashed-purple line), ii) large $l_s$ compared to $\delta$ still gives accurate results, iii) smaller $l_s$ compared to $\delta$ overestimates the effect of disorder variability, especially at lower porosities. However, at higher porosities the inaccuracy decreases independent of the choice of $l_s$.

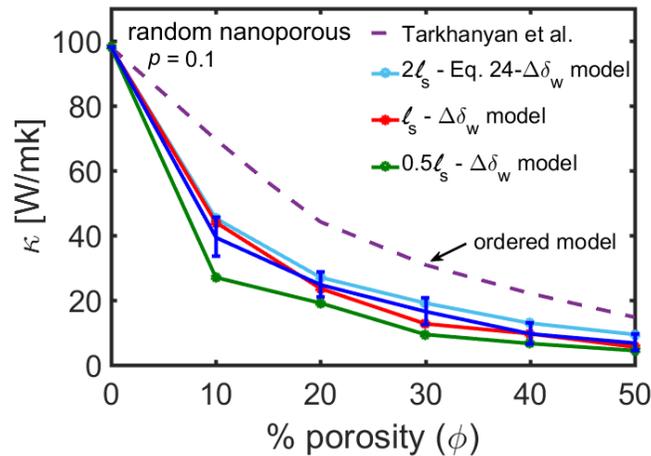

**Figure A6:** The sensitivity of the randomized models in the $l_s$ distance that we choose to split the channel into for the calculation of the porosity variation along the transport direction. Results for $2l_s$, $l_s$, and $l_s/2$, where $l_s=\delta$ are shown and compared to the Monte Carlo results and the simpler non-randomized model of Eq. 18.